\documentclass[useAMS,usenatbib]{mn2e}

\usepackage{times}
\usepackage{graphicx}
\usepackage{rotating}

\def\pcm3{{\rm\thinspace cm^{-3}}}

\def\n_h{{\rm n_{H}}}

\def\NH1{{$N_{\rm HI}~$}}

\def\ga{{\rm\thinspace gauss}}

\def\Msun{\hbox{$\rm\thinspace M_{\odot}$}}

\def\approxlt{\mathrel{\hbox{\rlap{\lower .5ex \hbox {$\sim$}}
        \raise .15 ex \hbox{$<$}}}}
\def\approxgt{\mathrel{\hbox{\rlap{\lower .5ex \hbox {$\sim$}}
        \raise .15 ex \hbox{$>$}}}}

\def\la{\mathrel{\hbox{\rlap{\hbox{\lower4pt\hbox{$\sim$}}}\hbox{$<$}}}}
\def\ga{\mathrel{\hbox{\rlap{\hbox{\lower4pt\hbox{$\sim$}}}\hbox{$>$}}}}
\newbox\grsign \setbox\grsign=\hbox{$>$} \newdimen\grdimen
\grdimen=\ht\grsign
\newbox\simlessbox \newbox\simgreatbox \newbox\simpropbox
\setbox\simgreatbox=\hbox{\raise.5ex\hbox{$>$}\llap
     {\lower.5ex\hbox{$\sim$}}}\ht1=\grdimen\dp1=0pt
\setbox\simlessbox=\hbox{\raise.5ex\hbox{$<$}\llap
     {\lower.5ex\hbox{$\sim$}}}\ht2=\grdimen\dp2=0pt
\setbox\simpropbox=\hbox{\raise.5ex\hbox{$\propto$}\llap
     {\lower.5ex\hbox{$\sim$}}}\ht2=\grdimen\dp2=0pt

\title[Blanco 1 brown dwarfs]{A photometric and astrometric investigation of the brown dwarfs in Blanco 1}
\author[ S. L. Casewell et al.]{ S. L. Casewell$^{1}$\thanks{E-mail:
slc25@le.ac.uk},  D .E. A. Baker$^{1,2}$, R. F. Jameson$^{1}$,  S. T. Hodgkin$^{3}$, P. D. Dobbie$^{4}$,
\newauthor and E. Moraux $^{5}$\\
$^{1}$Department of Physics and Astronomy, University of Leicester, University Road, Leicester LE1 7RH, UK\\
$^{2}$Science \& Technology Research Institute, University of Hertfordshire, College Lane, Hatfield AL10 9AB,UK\\
$^{3}$CASU, Institute of Astronomy,University of Cambridge, Maddingley Road, Cambridge, CB3 0HA, UK \\
$^{4}$University of Tasmania, Private Bag 50, Hobart, Tasmania 7001, Australia\\
$^{5}$UJF-Grenoble 1 / CNRS-INSU, Institut de Plan\'{e}tologie et d'Astrophysique 
de Grenoble (IPAG) UMR 5274, Grenoble, F-38041, France\\}

\begin{document}

\date{\today}

\pagerange{\pageref{firstpage}--\pageref{lastpage}} \pubyear{2011}

\maketitle

\label{firstpage}

\begin{abstract}

We present the results of a photometric and astrometric study of the low mass stellar and substellar population of the young open cluster Blanco 1. 
We have exploited $J$ band data, obtained recently with the Wide Field Camera (WFCAM) on the United Kingdom InfraRed Telescope (UKIRT), and 10 year old 
$I$ and $z$ band optical imaging from CFH12k and Canada France Hawaii Telescope (CFHT), to identify 44 candidate low mass stellar and substellar members, 
in an area of 2 sq. degrees, on the basis of their colours and proper motions. This sample includes five sources which are newly discovered. We also confirm
 the lowest mass candidate member of Blanco 1 unearthed so far (29M$_{\rm Jup}$). We determine the cluster mass function to have a slope of $\alpha$=+0.93,
 assuming it to have a power law form. This is high, but nearly consistent with previous studies of the cluster (to within the errors), and also that of its
 much better studied northern hemisphere analogue, the Pleiades.

\end{abstract}


\section{Introduction}

Open clusters are often acclaimed as excellent laboratories with which to study star formation. This is due to the co-eval nature of their members and estimates of their age being comparatively robust.  
Many open star clusters have been studied to date, yielding a large number of low mass members (e.g. \citealt{Baker_2010,Casewell_2007,Lodieu_2007a}) which have been used to refine our knowledge about the low mass end of star formation via mapping the initial mass function (IMF). The IMF, the number of objects per 
unit mass interval, is an observable outcome of star formation and can be used to critically examine theoretical models of this process.The IMF is commonly measured using an $\alpha$ parameter, where dN/dM $\propto$ M$^{-\alpha}$ and N is number of objects, and M is mass.
For most open star clusters (ages ~100 Myr), $\alpha$ is roughly consistent across all samples and $\approx$0.6 \citep{bouvier05}. This value is also consistent with.
field values such as those of \citet{Chabrier_2003}, although recently it has been suggested that for very low mass field brown dwarfs the IMF may have a different form. Indeed 
\citet{burningham10} suggest that in this case $\alpha$ may even have a negative value.

 In recent years there has been a particular 
emphasis on building a solid comprehension of the mechanisms by which very low mass  brown 
dwarfs and free-floating planetary mass objects form (e.g. \citealt{bate11}).
Nevertheless, key questions remain to be answered e.g. what is the lowest possible mass of object 
that can be manufactured by the star formation process$?$ From a theoretical stance, traditional models
predict that if substellar objects form like stars, via the fragmentation and collapse of molecular clouds, 
then there is a strict lower mass limit to their manufacture of 0.007-0.010 $\Msun$. This is set by the 
rate at which the gas can radiate away the heat released by the compression (e.g. \citealt*{low76}). However,
 in more elaborate theories, hypothetical magnetically mediated rebounds in collapsing 
cloud cores might lead to the decompressional cooling of the primordial gas, a lowering of the Jeans mass 
and hence the production of gravitationally bound fragments with masses of only $\sim$0.001 $\Msun$ \citep{Boss_2001}. 

However, while many surveys of open star clusters have been performed to search for substellar members, the majority of these are in the heavily populated Northern hemisphere clusters.
The lack of southern coverage from surveys (e.g. Sloan Digital Sky Survey \citealt{York_2000}; UKIRT Infrared Deep Sky Survey \citealt{Warren_2007}) has impeded detailed studies of the substellar population of a plethora of potentially interesting southern open clusters.

Blanco 1 is a 90$\pm$25 Myr \citep{Panagi_1997}  open cluster with an age similar to that of the 125 Myr Pleaides cluster \citep*{Stauffer_1998} at a distance of 207$\pm$12 pc as determined from $Hipparcos$ measurements \citep{Leeuwen_2009}. Recent work on the cluster includes spectroscopy of F and G type
stars \citep{Ford_2005} which  show that the metallicity is [Fe/H]=+0.04, with subsolar abundances for [Ni/Fe],
[Si/Fe], [Mg/Fe], and [Ca/Fe]. \citet*{cargile10} have determined a Lithium age for the cluster of 132$\pm$24 Myr which is closer to the age of the Pleiades than that of \citet{Panagi_1997}. We have taken the age of the cluster to be 120 Myr which is close to both measured values, and is present in the \citet{Chabrier_2000} DUSTY models.  

Recently \citet{platais11} surveyed 11 square degrees of the cluster to provide a comprehensive proper motion catalogue for all stellar objects down to M5V. \citet{Moraux_2007} performed the first study of the cluster to search for brown dwarfs using CFH12k on the Canada-France-Hawaii Telescope in the
 optical $z$ and $I$ bands to image 2.3 square degrees of the cluster centre.
They discovered $\approx$ 300 cluster members; 30-40 were estimated to be brown dwarfs, some of which had additional $K$ band photometry and
optical spectroscopy. Three of these objects were subsequently confirmed as members by \citet{cargile10}.
 
We have used the $I$ and $z$ band images from \citet{Moraux_2007} and have combined them with additional deep ($J\approx22$) $J$ band photometry obtained using WFCAM on UKIRT allowing us to not only select
fainter candidate cluster members, but also to measure the proper motion for some of the previously identified objects to prove if they are indeed associated with the cluster.

\section{Observations and  Data Reduction }
\subsection{CFHT12K data}
\begin{figure}
\begin{center}
\includegraphics[width=\columnwidth]{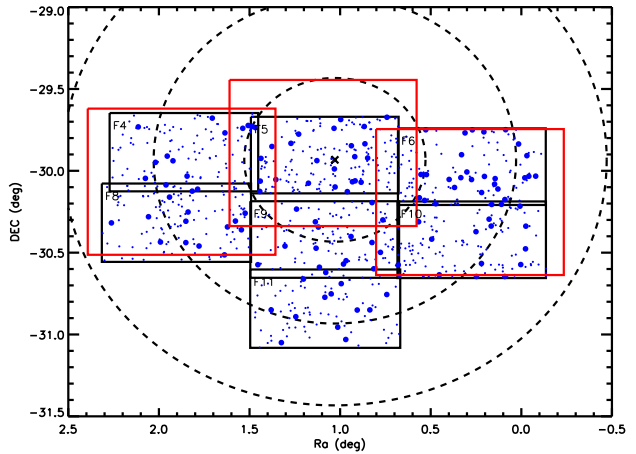}
\caption{Outline of the sky coverage of Blanco 1 from the CFHT12k tiles (black) and the WFCAM tiles (red), the low-mass and very-low-mass candidate lists of \citet{Moraux_2007} are shown as the small and large blue dots respectively. The black cross indicates the cluster centre, with circles of radius 0.5, 1.0 and 1.5 degrees being shown by the dashed black lines. }
\label{Figure_Blanco_CFHT_Coverage}
\end{center}
\end{figure}

The initial Blanco 1 data was taken with the CFHT12k optical mosaic 
camera during two separate runs as detailed in \citet{Moraux_2007}. The first of the runs 
occurred between 30 September 1999 and 2 October 1999, with the second occurring between 18 
and 20 of December 2000. A total of 7 fields were observed covering an area of 2.3 square 
degrees, in a (mostly) non overlapping pattern. Each separate field covered 
an area of 28'$\times$42'. The area of sky covered is shown in Figure~\ref{Figure_Blanco_CFHT_Coverage}. 
For each filter, Mould $I$ and $z$ Prime (see http://www.cfht.hawaii.edu/Instruments/Filters/cfh12k.html for filter profiles), a short observation of 10 s was accompanied by two longer 
600 s exposures. These were then combined to produce an equivalent image containing 1200s worth 
of exposure. The detection limits of the data were $I$$\sim$$z$$\sim$24 \citep{Moraux_2007},well below the stellar/substellar boundary which for Blanco 1 is estimated to lie at  $I\approx$19.15. The reduction of the initial data by \citet{Moraux_2007} 
followed the same prescription as described in \citet{Moraux_2003}.

The raw CFHT12k data frames were 
extracted from the Canadian Astrophysical Data Centre (CADC) archive and were re-reduced using the  imaging pipeline (\citealt*{Irwin_2001}) following the procedures described in \citet{Casewell_2007}.
Subsequently the two 600s images in each filter at each pointing were co-added prior to source extraction and  catalogue 
generation. Sources were identified as having a minimum of 5 interconnected pixels sitting 
at a significance of 1.5$\sigma$ above the background, with aperture photometry carried out 
using a radius of 3.5 pixels. In addition a morphological classification flag was provided 
with -1 indicating a stellar like profile, 0 noise and +1 non-stellar like sources. For the 
field-filter-extension/chip combinations of F4-$I$-10, F4-$z$-10, F6-$I$-6, F6-$z$-6 and F8-$I$-8 the astrometry 
needed further correction to that supplied by the CASU pipeline which was accomplished by using ``AAA'' rated stars in 2MASS. 

To refine the photometric calibration used by \citet{Moraux_2003} which was based on A0 stars, we calculated  a zero point for each chip in each filter using data from ESO. 

\begin{figure}
\begin{center}
\includegraphics[width=\columnwidth]{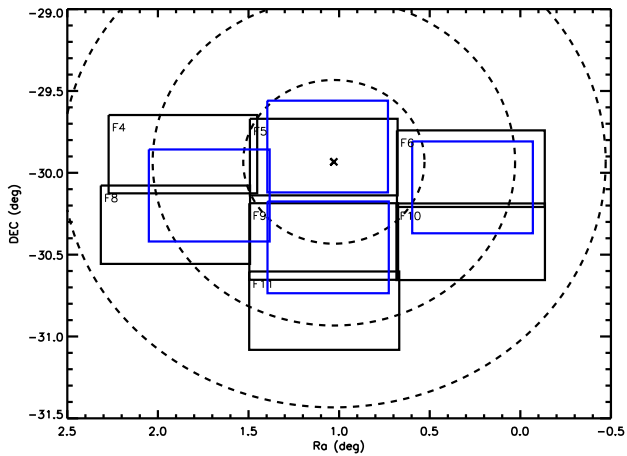}
\caption{Outline of the coverage of Blanco 1 offered by the Monitor project data compared to the coverage given by the original CFHT12k fields.}
\label{Figure_Blanco_Monitor_Coverage}
\end{center}
\end{figure}

 Blanco~1 formed part of a study of young open clusters by the Monitor project (e.g. \citealt{irwin08,irwin09} and references therein). The observations were obtained using the 
MPG/ESO 2.2-m telescope with WFI in service mode, with around 500 epochs measured between July 2005 and October 2007 for four pointings (See Figure~\ref{Figure_Blanco_Monitor_Coverage}). The instrument
 provides a field of view  of $\sim$34$\times$33 arcmin$^2$ (0.31 deg$^2$), using a mosaic of eight 2k$\times$4k pixel CCDs, at a scale of $\sim$0.238 arcsec pix$^{-1}$. The filter used was 
the ESO WFI broadband I filter (designated BB\# I203\_ESO879, also known as the I$_{EIS}$ filter) with a central wavelength of 826.9nm, and a sharp cutoff in the red shortwards of 950nm.

For a full description of the data reduction steps, the reader is referred to \citet{irwin2007}. Briefly, we used the pipeline for the INT wide-field survey \citep*{Irwin_2001} 
for 2D instrumental signature removal (bias correction, flat-fielding, defringing) and astrometric and photometric calibration. We then generated a master catalogue for each filter 
by stacking 20 of the frames taken in the best conditions (seeing, sky brightness and transparency) and running the source detection software on the stacked image. Astrometric 
calibration is tied into 2MASS and has residuals of better than 0.05 arcseconds per pointing. Photometric calibration of our data was carried out using regular observations of
 \citet{landolt92} equatorial standard star fields, measured as part of the standard ESO nightly calibrations. Prior to applying the calibration, we converted the Landolt (Cousins) 
photometry into the I$_{EIS}$ system using colour equations from Mike Irwin (private communication), see Equation~\ref{eqct} 
\begin{equation}
I_{EIS} = I - 0.03 (V-I)
\label{eqct}
\end{equation}(and also \citealt{irwin08}). By converting the 
Landolt standards into the $EIS$ system, and working entirely in the natural system of the instrument, this stage of the calibration of the CCD photometry is independent of 
the differences in colour between the cluster members which are somewhat redder than the Landolt standards.

The $I$ and $z$ zeropoints used to calibrate the CFHT data were then calculated by comparing the uncalibrated instrumental magnitudes against those from the Monitor project.
 Data from each of the 12 CFHT12k 
CCDs were binned for each of the two separate runs, i.e. all the objects found on chip 6 (over the different fields) taken in the 1999 run were combined to provide 
one single photometric zero-point for that chip. As small regions of 
overlap exist between some of the CFHT12k fields, a test of photometric accuracy was conducted for those objects with duplicate detections. This yielded 
RMS values of $\approx$0.035 and $\approx$0.040 for the $z$ and $I$ filters respectively. 

To confirm the calibration we first attempted to use the APASS survey (http://www.aavso.org/apass), however, there was insufficient overlap in magnitudes (the survey objects are saturated in the CFHT image) for us to use these data. We then cross-correlated the objects from \citet{Moraux_2007} with our data and used them to check the calibration. We obtain an offset of $I$=+0.066$\pm$0.018 and $z$=+0.080$\pm$0.012 between the original CFHT data and our reprocessed images, with the original data being fainter. This is marginally larger than the RMS scatter, but is in general smaller than the errors on the measurements, and so we are satisfied that our calibration is accurate. 

Following the photometric calibration, the separate $I$ and $z$ catalogues for each CFHT12k CCD chip were merged. 
This was done by using a flux limited sample of objects that had been morphologically classified as stellar. This subset was 
used as an input for pattern matching and linear transformation equation generation between the associated $x$ and $y$ 
pixel coordinates of the objects. Once a transformation had been established for the ``clean'' sample it was used to 
match the full sample together helping to reduce the number of spurious detections between the two images.

\subsection{WFCAM data}

In addition to the optical data, near-IR observation were also taken. Three WFCAM \citep{Casali_2007} $J$ band tiles were obtained in UKIRT service mode, 2 on the night of 31 October 2006 and 1 on the night of 22 July 2009. Each WFCAM pawprint 
used exposures of 18 s and a 9 point jitter pattern with 2x2 microstepping to improve the spatial sampling, making  1 hour of observations in total per tile:600s exposure per pawprint, in seeing of $\approx$ 1'' or better. The WFCAM data was processed as for the Pleiades survey of \citet{Casewell_2007}. The calibration and pipeline for the data reduction are described in \citet{hodgkin09}.The photometric calibration is tied to 2MASS photometry resulting in accuracies of $\sim$1.5 per cent.  The total area covered by the WFCAM fields is 2.25 square degrees, of which $\approx$2 square degrees overlaps with the CFH12k data as shown in 
Figure~\ref{Figure_Blanco_CFHT_Coverage}.

The CFHT12k data was pattern matched to the WFCAM data on an individual field by field, chip by chip basis to minimise multiple detections. 
Each source also had to be classified as stellar in both the CFHT12k $I$ and $z$ band images as well as the WFCAM $J$ band image. The resulting catalogue 
contained 9853 sources (8440 of which were unique).

We estimated the completeness of both the CFHT12k and WFCAM images using the method described in \citet{Casewell_2007}. We inserted 200 fake stars (12$<J<$22, 15$<I,z<$30) generated by \textsc{iraf} 
into each chip, 10 times to enable us to have sufficient objects on which to perform the statistics. The skylevel, detector gain, seeing, exposure times and zeropoint of the images was taken into 
account when creating the fake stars. The CASU routine \textsc{imcore} was used to extract the objects from each image and then the numbers of inserted and extracted objects were compared per 
magnitude bin. The data were found on average to be 90 per cent complete at 19.73 in the $J$ band, and 21.5 and 20.6 in the $I$ and $z$ bands respectively, and 50 per cent complete at 20.8 in 
$J$ and  22.2 and 21.5 in the $I$ and $z$ bands respectively. In general it was found that chip number 4 of WFCAM was about 0.2 mags less sensitive than the other 3 chips. 

\section{Results}
\subsection{Photometric selection}

\begin{figure*}
\begin{center}
\scalebox{0.5}{\includegraphics[angle=270]{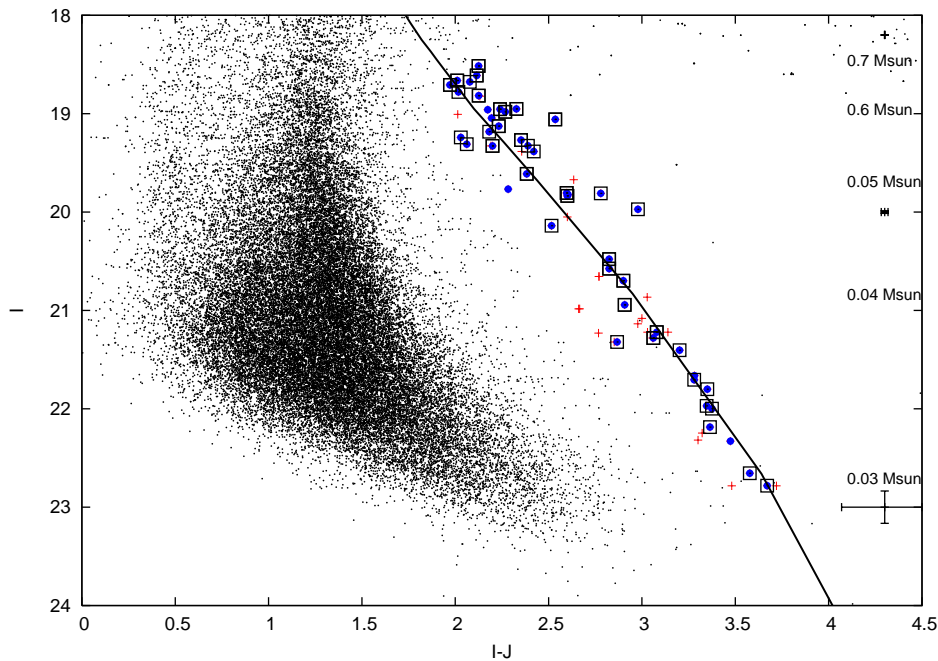}}
\caption{Colour magnitude selections in  $I$, $I-J$. The black points are the stellar CFHT12k-WFCAM sources. 
Red + mark the selected objects, while large blue filled circles indicate objects that remained after the proper motion selection.
The objects identified as candidate members from \citet{Moraux_2007} are marked by boxes. Representative errorbars and masses are also shown, as is the DUSTY model isochrone for 120 Myr \citet{Chabrier_2000}.}
\label{Figure_Blanco_Colour_Cuts}
\end{center}
\end{figure*}

For consistency with previous studies of objects within this effective temperature range we chose to use the DUSTY models of \citet{Chabrier_2000}. We selected all sources with $I$-$J$ within 0.5 of each side of the model for 120 Myrs at 207 pc. This selection allows for uncertainty in distance and the equal mass binary sequence.  We then applied additional selection criteria of $I-J>$1.95 and $z-J>1.15$ to extract the sequence from
the bulk of the field stars. Figure \ref{Figure_Blanco_Colour_Cuts}. We selected a total of 83 objects using this method.

To determine the accuracy of our selection criteria, we compared how many of the objects presented in \citet{Moraux_2007} were recovered by our survey.
\citet{Moraux_2007} present 764 unique sources, titled Low Mass Candidates (LMC) and Very Low Mass Candidates (VLMC), 578 of which are located within our survey area. Many of these objects were discovered using short exposures of 10s and so are 
saturated in our data (1200 s in the $I$ and $z$ bands). To allow for this, and to better exploit our deeper data, we then applied a bright limit of $I=18.5$ in our selection criteria. 
We recovered 522 objects in our survey area before the selection criteria were applied. The missing $\approx$ 50 objects have not been recovered due to falling between chip gaps, or not being detected in our $J$ band data as they are not red enough to be cluster members.
Of the very low mass candidates, 81 are covered by our survey, and we recover all of these objects in our data.  

However, after the selection cuts were made,only 27 LMC objects remained and 38 VLMC objects although there is some overlap between the lists (Table \ref{estelle}; Appendix A). The remainder were lost as they were brighter than $I=18.5$, or fell outside the strip defined by the model i.e. despite 
appearing to belong to the cluster sequence in $I-z$, they do not appear to belong to the sequence in $I-J$ or $z-J$ (generally being too blue), and so are probably not members of the cluster.

\subsection{Proper motions and membership probabilities}
We measured the  proper motions of the 83 selected objects using the $z$ and $J$ bands, which gave an  epoch difference of 10 years for the majority of objects, although a handful had a shorter epoch difference of only 8 years.
The $z$ band was chosen over the $I$ band to minimise any effects of differential chromatic refraction as all images were taken at high airmass, due to Blanco 1 being 
near the observing limits of both UKIRT and CFHT.
We used a  pixel-pixel transformation routine that uses a set of 
stationary reference sources in each image as described by \citet{Casewell_2007}. 
The reference objects 
were selected to have magnitudes 16 $\leq$ mag $<$ 20 in  $z$, similar to that 
of the candidates, but not so faint as to have poor pixel centroiding. It was also required that they have 
an ellipticity of less than 0.2 in the $z$ band image and be located within 10' of the candidate to minimise radial distortion effects.In regions of overlap it was also ensured that the candidate be on the
same chip as the reference stars in each image.

Centroiding errors were estimated using fake stars as for the completeness calculations, only this time the difference in pixel positions between the inserted and recovered stars was measured. 
This difference was measured in magnitude bins, as it was anticipated that fainter objects would have larger centroiding errors. These errors were 0.01 pixels in $J$ for $J<17.0$ and 0.07 for $J>17.0$, where the detector pixel size is 0.4''.
For the $z$ band, the errors were 0.03 for $z<23.0$, where the detector pixel size is 0.2''. These pixel measurements were added quadratically to the rms error on the pixel-pixel transforms to generate the proper motion errors.

Once we had measured the proper motions, the data were then binned  in 10 mas yr$^{-1}$ bins in both RA and dec, and a 2D Gaussian was fitted to the data in proper motion space.  The $\sigma$ derived was then used to reject objects 
outside the 2$\sigma$ boundary to remove outliers, and the fit was then recalculated. This gave a Gaussian width of $\sigma$$\sim$9.0 mas yr$^{-1}$ to be used for candidate selection.

\begin{figure}
\begin{center}
\scalebox{0.45}{\includegraphics[angle=270]{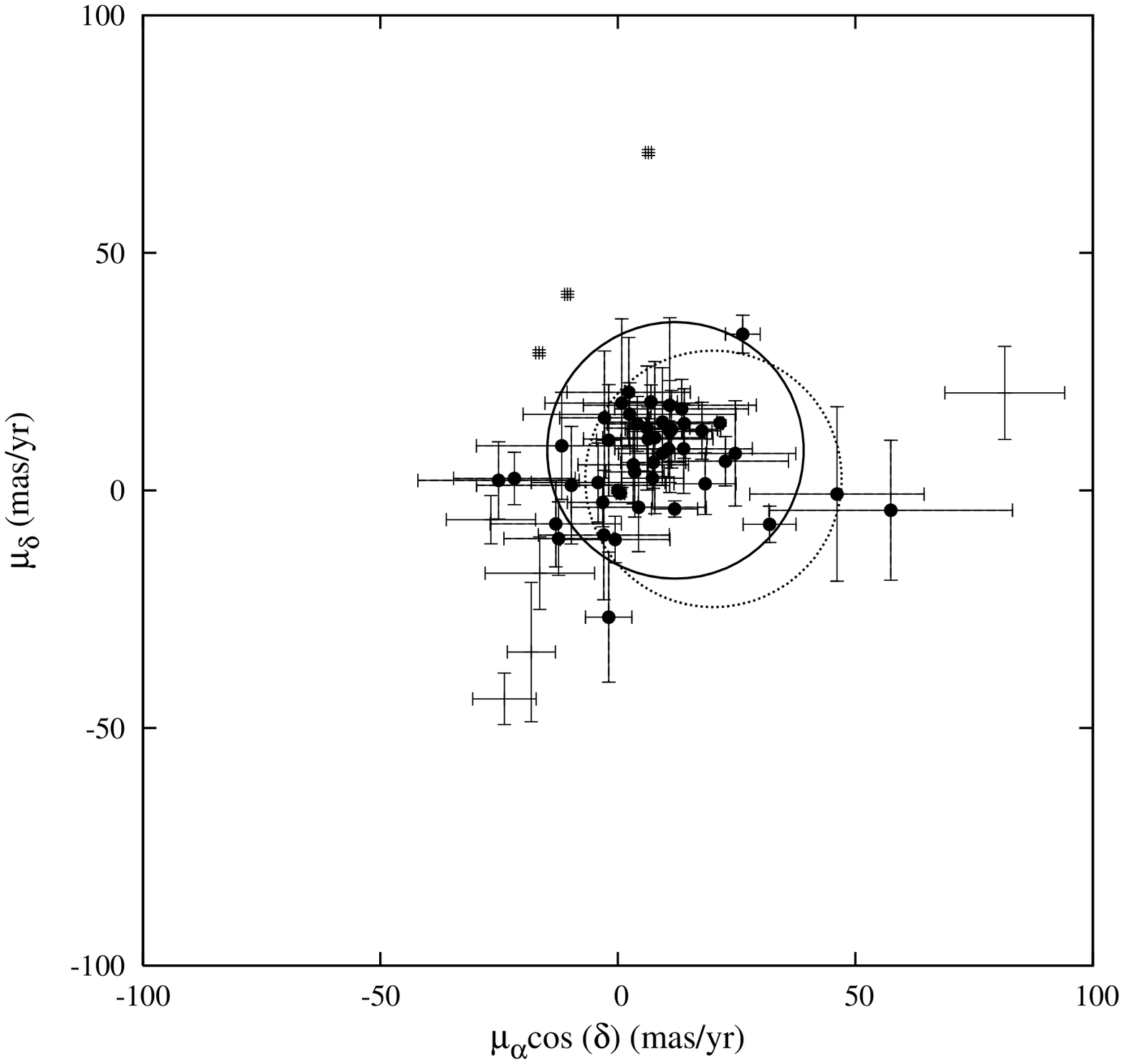}}
\caption{Proper motion diagram for the Blanco 1 cluster.  Objects marked by a + are all objects we measured proper motions for. Objects marked by a filled circle are selected objects; they fall within a 27 mas yr$^{-1}$
circle centred on the cluster proper motion once it has been adjusted for the field star motions ($\mu_{\alpha\cos\delta}=12.11$ mas yr$^{-1}$, $\mu_{\delta}=8.43 $mas yr$^{-1}$).
 The centre of the dotted circle shows the unshifted cluster proper motion ($\mu_{\alpha\cos\delta}$ = 20.11 mas yr$^{-1}$, $\mu_{\delta}$=2.43 mas yr$^{-1}$).}
\label{Figure_Blanco_1_PM}
\end{center}
\end{figure}

It is  obvious from Figure \ref{Figure_Blanco_1_PM} that the average proper motion of our selected objects ($\mu_{\alpha\cos\delta}$ = 8.93 mas yr$^{-1}$, $\mu_{\delta}$=6.70 mas yr$^{-1}$) is significantly different from the 
literature value of the cluster proper motion ($\mu_{\alpha\cos\delta}$ = 20.11 mas yr$^{-1}$, $\mu_{\delta}$=2.43 mas yr$^{-1}$; \citealt{Leeuwen_2009}). \citet{platais11} determined that the mean motion of field stars is not at 0,0 as is generally used for relative proper motions, but at  $\mu_{\alpha\cos\delta}$ = 8.0 mas yr$^{-1}$, $\mu_{\delta}$=-6.0 mas yr$^{-1}$. To determine if this offset was applicable to our data we modified our photometric selection criteria to obtain everything 0.5 mag bluer in $I-J$ than the Dusty model \citep{Chabrier_2000}. We then measured proper motions for these 250 objects, and fitted a 2D Gaussian to their proper motions as before. We determined that the centre of this distribution is at $\mu_{\alpha\cos\delta}$ = 1.96 mas yr$^{-1}$, $\mu_{\delta}$=-3.64 mas yr$^{-1}$, with a width of 12 mas yr$^{-1}$. This motion is smaller than that measured by \citet{platais11}, but they measured many more stars, and to a better accuracy than this work which has concentrated on the fainter members of the cluster.  This mean motion explains the offset between our candidate distribution and the cluster motion. Taking into account the offset makes our mean proper motion $\mu_{\alpha\cos\delta}$ = 16.93 mas yr$^{-1}$, $\mu_{\delta}$=0.70 mas yr$^{-1}$ which is much closer to the reported value for the cluster. We used the literature value of the cluster proper motion, minus the \citet{platais11} estimation of the field star motion, $\mu_{\alpha\cos\delta}=12.11$ mas yr$^{-1}$, $\mu_{\delta}=8.43 $mas yr$^{-1}$, as the cluster centre for selection purposes.  It should also be noted that the field and cluster stars are not that far apart in terms of the relative errors which \citet{platais11} discuss in more depth in their work on Blanco 1. Despite the small difference between the proper motion of the cluster and field stars, the narrower dispersion in proper motion for the cluster objects, and the colour selections in $I-J$ and $z-J$ mean we can be confident that we are selecting true cluster members. 

Of the 83 candidates for which we obtained astrometry, 44 had proper motions within (or with errors within) 3$\sigma$ of the cluster value adjusted to take into 
account the field star relative motion ($\mu_{\alpha\cos\delta}=12.11\pm0.38$, $\mu_{\delta}=8.43\pm0.25$; \citealt{platais11,Leeuwen_2009}).
Of these 44 members, 33  are present in the VLMC list and 24 are present on the LMC list, with 18 objects common to both lists of candidate sets. This leaves 5 new low-mass candidate members to the cluster. The previously identified objects that were rejected are LMC694/VLMC66,  VLMC64, VLMC68, VLMC69, VLMC71 and VLMC74.  All VLMC objects with spectra remain in our candidates apart from objects 29,38,48 and 49 which are not in our survey area, and 16 and 22 which did not meet our photometric selection criteria (they are too bright). It should be noted that the 3 objects identified by \citep{Moraux_2007} as non-members based upon their spectroscopy, VLMC28, 37 and 44 were not selected, (VLMC28 also does not have a spectrum indicative of it being a cluster member; \citealt{cargile10}). The sucess of this method in recovering the previously identified spectroscopic members, despite large errors on the small cluster proper motion, leads us to believe this is a robust method for determining cluster members.

Examining the proper motion vector point diagram (Figure~\ref{Figure_Blanco_1_PM}) it is clear 
that towards the location of the cluster there is an overdensity of objects when compared with regions at a similar distance from 0,0
but on the opposing side of the field star distribution. Unfortunately, the low number of sources coupled with a cluster proper motion 
comparable to the average proper motion error means that the two Gaussian approach to calculating membership probability as used by \citet{Baker_2010} is not applicable. Instead the simpler 
annulus method was used as in \citet{Casewell_2007}. All non-selected objects with measured proper motions (barring those where there has obviously been a problem with the fit - 7 cases) 
were used per $I$  magnitude bin to assess the contamination and membership probabilities. We were unable to use an annulus centred on 0,0 as there are very few objects there not deemed to be members of Blanco 1 due to the low proper motion of the cluster.
This may mean that some of the membership probabilities have been underestimated (Table \ref{prob}).


\begin{table}
\caption{\label{prob}Magnitude bins and the associated membership probability and completeness at this $I$ magnitude.}
\begin{center}
\begin{tabular}{lcc}
\hline
$I$&Probability& Completeness\\
\hline
$<$19.5 &91\%&100\%\\
19.5-20.5&75\%&97\%\\
20.5 to 21.5&50\%&93\%\\
21.5-22.5&50\%&62\%\\
$>$22.5&40\%&16\%\\
\hline
\end{tabular}
\end{center}
\end{table}

 \begin{table*}
\caption{\label{table}Name, \citet{Moraux_2007} name,coordinates,proper motion, $I$, $z$, $J$ magnitudes for our members to the cluster. Previously discovered 
members also also have their other known names
listed. An atserisk (*) indicates that membership has been confirmed from \citet{Moraux_2007} spectroscopy.}
\begin{center}
\begin{tabular}{l c c c c c c c c }
\hline
Name &Alternate name&RA &dec&$\mu_{\alpha}$cos$\delta$&$\mu_{\delta}$&$I$&$z$&$J$\\
&&\multicolumn{2}{|c|}{J2000.0}&\multicolumn{2}{|c|}{mas yr$^{-1}$}&&&\\
\hline
bl2399-4716  &   & 00   00 6.83  &-30   13 33.72&	+2.27 $\pm$12.96 &+20.66  $\pm$11.56&	19.045$\pm$0.010&18.201$\pm$0.033&16.851$\pm$0.007\\
bl28626-47167 &LMC571/VLMC19  &00   07 50.63  &-30    5  9.97&	+7.45 $\pm$6.86 &+5.90  $\pm$5.45&	18.515$\pm$0.013&17.688$\pm$0.037&16.391$\pm$0.005\\
bl2868-4699  &VLMC89   &00   00 5.86  &-30   20 18.39&	+13.99 $\pm$8.10 &+14.07  $\pm$7.33&	21.971$\pm$0.050&20.755$\pm$0.073&18.625$\pm$0.027\\
bl28691-43204 &VLMC54   &00    07 41.46  &-29   56 20.39&	+31.93 $\pm$5.57 &-7.13  $\pm$3.81&	19.971$\pm$0.017&18.782$\pm$0.041&16.993$\pm$0.008\\
bl32426-33057&LMC629/VLMC46   &00    05 57.04  &-29   43 48.33&	-1.93 $\pm$4.90 &-26.66  $\pm$13.69&	19.328$\pm$0.015&18.431$\pm$0.039&17.128$\pm$0.009\\
bl32697-33301  &LMC729/VLMC72 &00    06  8.96  &-29   44 25.22&	+2.45 $\pm$22.41 &+16.06  $\pm$6.60&	21.280$\pm$0.029&20.035$\pm$0.054&18.219$\pm$0.019\\
bl33600-62347 &LMC626  &00    06 41.73  &-29   42 52.52&	-13.07 $\pm$13.81 &-7.03  $\pm$9.07&	19.310$\pm$0.015&18.466$\pm$0.039&17.248$\pm$0.009\\
bl33721-62566  &LMC705/VLMC63 &00    06 49.36  &-29   40 41.18&	+10.90 $\pm$8.92 &+12.40  $\pm$10.72&	20.698$\pm$0.022&19.521$\pm$0.046&17.798$\pm$0.013\\
bl34536-62053  &VLMC118 &00    06 32.59  &-29   46  05.57&	+11.29 $\pm$9.71 &+12.87  $\pm$8.17&	22.782$\pm$0.086&21.357$\pm$0.097&19.111$\pm$0.037\\
bl37640-49624  &LMC592/VLMC30 &00    08 27.38  &-29   43 54.26&	+10.94 $\pm$18.20 &+17.94  $\pm$18.40&	18.708$\pm$0.014&17.965$\pm$0.037&16.737$\pm$0.006\\
bl3800-15178    &LMC580/VLMC25*&00   00 42.74  &-30   17 43.43&	+6.40 $\pm$13.62 &+10.82  $\pm$6.75&	18.612$\pm$0.010&17.779$\pm$0.032&16.497$\pm$0.006\\
bl3819-15319   &LMC621 &00   00 36.44  &-30   19 15.95&	+22.62 $\pm$13.28 &+6.14  $\pm$5.19&	19.184$\pm$0.011&18.288$\pm$0.033&17.002$\pm$0.008\\
bl43328-55651  &VLMC57 &00    07 22.76  &-30    01 57.32&	-4.16 $\pm$14.08 &+1.65  $\pm$8.34&	20.139$\pm$0.018&18.952$\pm$0.043&17.623$\pm$0.012\\
bl44742-27543  &VLMC82 &00    04 42.00  &-30    04 33.52&	+11.93 $\pm$4.88 &-3.91  $\pm$1.70&	21.704$\pm$0.041&20.497$\pm$0.058&18.425$\pm$0.034\\
bl46456-37665   &VLMC70&00    05 45.83  &-30    03 46.14&	+24.71 $\pm$12.74 &+7.79  $\pm$11.09&	21.221$\pm$0.028&20.031$\pm$0.049&18.142$\pm$0.019\\
bl51473-23799   &LMC582/VLMC26&00    04 54.98  &-29   46 32.88&	+7.79 $\pm$10.99 &+11.10  $\pm$16.02&	18.663$\pm$0.010&17.935$\pm$0.031&16.653$\pm$0.006\\
bl51714-24041 &LMC719/VLMC67  &00    05  05.19  &-29   49 55.81&	+13.41 $\pm$14.08 &+17.17  $\pm$6.20&	20.943$\pm$0.025&19.925$\pm$0.042&18.035$\pm$0.016\\
bl54505-36470 &  &00    03 24.26  &-30   00 51.66&	+0.82 $\pm$16.17 &+18.36  $\pm$17.75&	18.959$\pm$0.010&18.094$\pm$0.031&16.785$\pm$0.007\\
bl54514-36502 &LMC608/VLMC41*  &00    03 23.62  &-29   55 17.56&	+0.00 $\pm$0.63 &-0.00  $\pm$0.63&	19.058$\pm$0.010&18.033$\pm$0.031&16.522$\pm$0.006\\
bl54613-36058  &LMC663/VLMC60 &00    03 40.17  &-30    03 40.85&	-0.57 $\pm$11.55 &-10.33  $\pm$4.90&	20.478$\pm$0.017&19.384$\pm$0.038&17.654$\pm$0.012\\
bl54805-35986  &VLMC74 &00    03 43.78  &-30    04  01.66&	+26.29 $\pm$3.66 &+32.89  $\pm$3.99&	21.404$\pm$0.030&20.167$\pm$0.047&18.202$\pm$0.019\\
bl55543-35398 &LMC645/VLMC51*  &00    04  07.62  &-29   59 18.78&	+3.57 $\pm$6.12 &+3.91  $\pm$9.53&	19.806$\pm$0.014&18.782$\pm$0.036&17.210$\pm$0.009\\
bl56159-35803  & &00    03 50.87  &-30    01 58.01&	+17.68 $\pm$3.98 &+12.55  $\pm$5.98&	22.329$\pm$0.070&21.084$\pm$0.087&18.855$\pm$0.032\\
bl56339-2865  &LMC624/VLMC45*  &00    01 35.71  &-30    03 10.13&	+18.36 $\pm$6.54 &+1.41  $\pm$6.49&	19.268$\pm$0.012&18.326$\pm$0.033&16.915$\pm$0.008\\
bl57823-3435 &VLMC114   &00    02  5.68  &-30    01 23.65&	+10.59 $\pm$8.078 &+8.75  $\pm$5.89&	22.655$\pm$0.110&21.382$\pm$0.105&19.077$\pm$0.042\\
bl57973-12146  &LMC647/VLMC55 &00    02 15.12  &-30    09 52.64&	-1.94 $\pm$2.57 &+10.56  $\pm$11.70&	19.835$\pm$0.016&18.843$\pm$0.036&17.235$\pm$0.009\\
bl58756-8496 &VLMC65   &00   00 7.31  &-29   54 26.81&	-2.81 $\pm$9.46 &+15.26  $\pm$14.08&	20.575$\pm$0.021&19.495$\pm$0.043&17.751$\pm$0.015\\
bl6053-16818 &VLMC53   &00    02 16.10  &-30   18 41.03&	+13.82 $\pm$14.46 &+8.77  $\pm$9.38&	19.810$\pm$0.013&18.747$\pm$0.034&17.031$\pm$0.008\\
bl60868-7134   &VLMC32 &00    01 19.28  &-29   54  06.58&	+57.44 $\pm$25.62 &-4.17  $\pm$14.72&	18.779$\pm$0.010&17.991$\pm$0.031&16.762$\pm$0.007\\
bl63549-4173 &LMC639/VLMC50*   &00   00 9.86  &-30    01 59.39&	-9.78 $\pm$19.95 &+1.10  $\pm$12.38&	19.612$\pm$0.013&18.651$\pm$0.034&17.229$\pm$0.011\\
bl65492-14209 &VLMC83  &00   00 34.79  &-30    02 51.04&	+21.52 $\pm$0.63 &+14.18  $\pm$0.63&	21.800$\pm$0.044&20.494$\pm$0.059&18.450$\pm$0.026\\
bl71077-57153  &LMC609/VLMC36* &00    07  08.79  &-30    06 42.35&	+7.29 $\pm$6.58 &+2.57  $\pm$6.10&	18.958$\pm$0.014&18.072$\pm$0.038&16.713$\pm$0.006\\
bl73194-41854 &LMC632  &00    08 13.55  &-30   16 50.17&	-12.48 $\pm$11.52 &-10.16  $\pm$7.71&	19.384$\pm$0.015&18.208$\pm$0.038&16.964$\pm$0.008\\
bl7507-10415  &LMC585  &00   00 12.10  &-30   35 56.07&	+6.15 $\pm$8.99 &+13.13  $\pm$13.04&	18.676$\pm$0.010&17.833$\pm$0.031&16.599$\pm$0.006\\
bl76145-40818  &LMC595/VLMC33 &00    06 10.79  &-30   21 37.02&	+3.23 $\pm$11.65 &+5.40  $\pm$8.18&	18.817$\pm$0.014&17.961$\pm$0.037&16.691$\pm$0.007\\
bl76187-40858  & &00    06  08.53  &-30   25 42.12&	-2.96 $\pm$13.80 &-9.37  $\pm$13.61&	19.767$\pm$0.017&18.851$\pm$0.040&17.483$\pm$0.011\\
bl76366-40509 &VLMC75  &00    06 29.32  &-30   20 33.54&	+4.38 $\pm$14.23 &-3.55  $\pm$9.32&	21.321$\pm$0.035&20.260$\pm$0.054&18.454$\pm$0.023\\
bl78357-42493 &LMC631  &00    07 33.84  &-30   24 35.13&	+9.32 $\pm$9.17 &+7.78  $\pm$6.73&	19.327$\pm$0.015&18.376$\pm$0.039&16.937$\pm$0.008\\
bl84053-20433 &  &00    03 17.87  &-30   11 40.93&	+6.94 $\pm$5.57 &+18.65  $\pm$3.55&	21.663$\pm$0.037&20.350$\pm$0.052&18.382$\pm$0.023\\
bl85709-22911  &LMC604/VLMC43* &00    04 32.88  &-30   18 41.90&	+9.41 $\pm$11.43 &+14.39  $\pm$11.46&	19.127$\pm$0.010&18.238$\pm$0.031&16.894$\pm$0.008\\
bl862-2266   &LMC600/VLMC34*   &00    01 48.76  &-30   38  06.81&	-3.19 $\pm$7.45 &-2.53  $\pm$5.11&	18.951$\pm$0.011&17.960$\pm$0.032&16.624$\pm$0.007\\
bl87593-22425 &VLMC93  &00    04 57.74  &-30   14  02.01&	+46.11 $\pm$18.36 &-0.75  $\pm$18.34&	22.186$\pm$0.073&20.903$\pm$0.058&18.822$\pm$0.034\\
bl89514-11054 &LMC619  &00    02 50.60  &-30   28 53.89&	-11.83 $\pm$17.94 &+9.38  $\pm$11.27&	19.241$\pm$0.011&18.429$\pm$0.033&17.211$\pm$0.010\\
bl90054-11175  &VLMC85 &00    03  06.63  &-30   29 53.90&	+4.08 $\pm$6.31 &+14.00  $\pm$5.78&	21.999$\pm$0.049&20.763$\pm$0.067&18.624$\pm$0.029\\
\hline
\end{tabular}
\end{center}
\end{table*}

\section{Mass Spectrum}

\begin{figure*}
\begin{center}
\scalebox{0.4}{\includegraphics[angle=270]{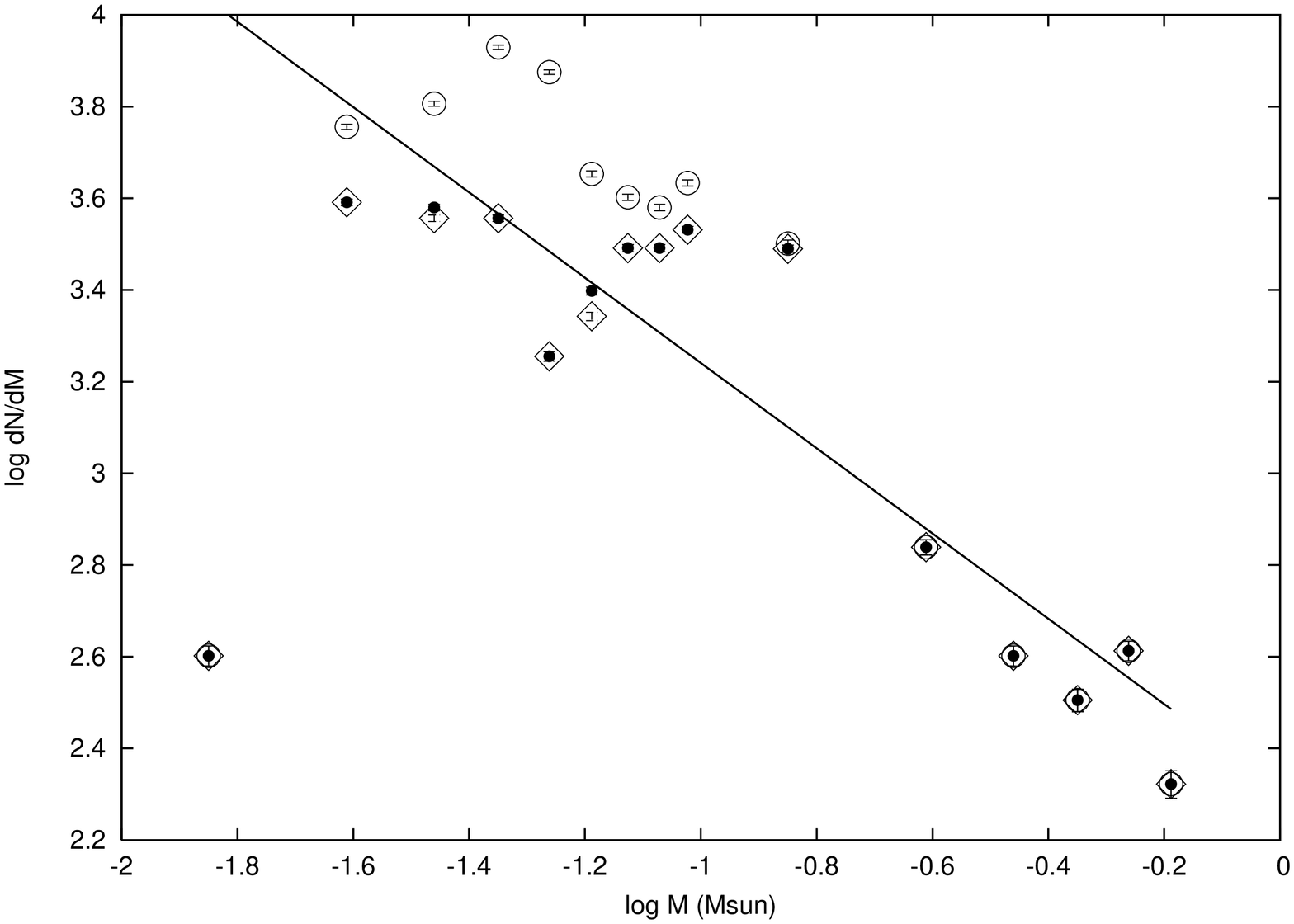}}
\caption{The mass spectrum for Blanco 1. The circles are the original members detailed by \citet{Moraux_2007}. It should be noted that the errors 
are poissonian, and in the case of the lowest mass bins, actually fall within the plotted point. The diamonds are the \citet{Moraux_2007} objects, 
with non-members as determined from this work, removed. The filled circles are a complete mass fucntion for all objects we determined to be members. 
The fit to the data of $\alpha$=0.93$\pm$0.11 is the solid line.}
\label{mass_func}
\end{center}
\end{figure*}

We were not able to generate  mass functions for our new candidate members alone, due to there being too few objects to be statistically significant. However, we have used the \citet{Moraux_2007} members combined with the $I-z$ colour and the NextGen model \citep{Baraffe_1998} for objects brighter than $I$=20.0, and the DUSTY model \citep{Chabrier_2000} for object fainter than this. Both models are for 120 Myr.  We present 3 mass spectra, one the original data from \citet{Moraux_2007}, and the second with the non-members, as determined from this work, excluded. The third dataset is all the members, including our 5 new objects (Figure \ref{mass_func}). These new objects have masses between 35 M$_{\rm Jup}$ and 46 M$_{\rm Jup}$, whereas the whole mass range of candidates is between 29 and 80 M$_{\rm Jup}$.
The third dataset (filled circles)is fitted by the straight line in Figure \ref{mass_func}, with an $\alpha$ value of 0.93$\pm$ 0.11. However, to obtain this fit, we have omitted 
the point at log M = -1.85: the faintest objects and lowest mass bin where we know the incompleteness is largest. We do not have a good enough photometry to accurately separate 
the single and binary star sequences, and thus suspected binaries have been assigned a mass equivalent to that of a single object at their recorded magnitude. The point at 
log M=-0.85 is discrepantly high and appears to be affected by binaries and possibly higher multiples at $I\approx$20 (Figure \ref{Figure_Blanco_Colour_Cuts}). 

 The $\alpha$=0.93$\pm$0.11 indicates the slope is higher, but is consistent (to within the errors) with the values given by \citet{Moraux_2007}, which are 0.67$\pm$0.14 
and 0.71$\pm$0.13 for 100 and 150 Myr models respectively, especially considering we have small number statistics in some mass bins.

While we have discovered 5 new low mass cluster members, our $J$ band data did not allow us to probe deeper into Blanco 1 than the original $Iz$ survey by \citet{Moraux_2007}. This situation will be improved as the VISTA VIKING survey which aims to cover the whole of the Blanco 1 cluster. It will also provide deeper multi-band photometry, as well as a far greater  baseline between observations to be used for proper motion analysis. Our work has shown that Blanco 1 does contain brown dwarfs as low in mass as $\sim$30 M$_{\rm Jup}$, making it similar to the Pleiades.

Follow up spectroscopic data  will allow us to place constraints on the binary fraction of the cluster, as well as confirm membership for the  objects without spectra. Decreasing the errors on the proper motions will allow us to determine members with much more confidence than we are currently able to due to the low space motion of the cluster. Once a full census of the cluster has been performed, Blanco 1 can then be properly compared to clusters such as the Pleiades. One can then test for the environmental tolerance of the IMF, dynamical evolution and mass segregation effects as well as providing 
further observational constraints to compare with the results of brown dwarf formation simulations.

\section{Summary}

We have used near-IR and optical photometry with proper motions derived from CFHT $z$ and WFCAM $J$ band images to identify 44 candidate cluster members with masses
 between 29 M$_{\rm Jup}$  and 80 M$_{\rm Jup}$. 5 of these are previously unidentified candidate members and 40 have been identified by \citealt{Moraux_2007}, 8 of
 which have been  confirmed as
 cluster brown dwarfs from  spectra. We derive $\alpha$=0.93$\pm$0.11 from the mass spectrum, which  is consistent with the literature for this cluster.

\section{Acknowledgements}

DEAB acknowledges support from STFC and SLC is supported by a  University of Leicester post-doctoral research position.  This paper made use of data from the United Kingdom Infrared Telescope which is operated by the Joint Astronomy Centre on behalf of the Science and Technology Facilities Council of the U.K.We would also like thank France Allard for the use of the Phoenix web simulator at http://phoenix.ens-lyon.fr/simulator/index.faces which has been used in this research.
This research has also made use of NASA's Astrophysics Data System Bibliographic Services.

\bibliographystyle{mn2e}

\bibliography{mnemonic,blanco1}
\newpage
\appendix
\section{Supplementary table}
Table \ref{estelle} displays a list of the Low Mass Candidates (LMC) and Very Low Mass Candidates (VLMC) identified by \citet{Moraux_2007}, that 
were present in our survey area, but were rejected by our photometric selection.

\begin{table}

\caption{\label{estelle}LMC and VLMC candidate members in our data, but determined to be non-members of the cluster due to photometry that was incompatible with
our selection criteria.}

\begin{center}

\begin{tabular}{l l}
\hline
Name& Name\\
\hline
LMC573 &LMC678 \\   
LMC574 &LMC679 \\   
LMC576 &LMC682 \\   
LMC578 &LMC683 \\   
LMC581/VLMC28&LMC684 \\   
LMC586 &LMC685 \\   
LMC587 &LMC686 \\   
LMC588 &LMC688 \\   
LMC590 &LMC689 \\   
LMC593 &LMC690 \\   
LMC594 &LMC692 \\   
LMC596 &LMC695 \\   
LMC597 &LMC699 \\   
LMC598 &LMC700 \\   
LMC603/VLMC39&LMC702 \\  
LMC610 &LMC704 \\   
LMC611 &LMC707 \\   
LMC614 &LMC708 \\   
LMC616 &LMC709 \\   
LMC620 &LMC710 \\    
LMC625 &LMC711 \\   
LMC627 &LMC712 \\   
LMC630 &LMC713 \\   
LMC635 &LMC715 \\   
LMC637 &LMC716 \\   
LMC638 &LMC717 \\   
LMC640 &LMC718 \\   
LMC642 &LMC720 \\   
LMC643 &LMC721 \\   
LMC644 &LMC722 \\   
LMC646 &LMC724 \\   
LMC652 &LMC725 \\   
LMC655/VLMC58&LMC726 \\
LMC656 &LMC727 \\   
LMC657 &LMC728 \\   
LMC658 &LMC730 \\   
LMC659 &LMC732 \\   
LMC660 &LMC735 \\   
LMC661 &LMC736 \\   
LMC662 &LMC738 \\   
LMC666 &LMC739 \\   
LMC668 &LMC741 \\   
LMC669 &LMC747 \\   
LMC671 &LMC749 \\   
LMC673 &LMC752 \\   
LMC674 &LMC755 \\   
LMC675 &LMC762 \\   
LMC677 &\\
\hline
\end{tabular}
\end{center}
\end{table}

\label{lastpage}

\end{document}